\documentclass{PoS}
\usepackage{amsmath,amssymb}
\graphicspath{{./Figures/}}

\newcommand{\preprint}{
  \begin{picture}(0,0)
    \put(232,130){{\rm\normalsize ADP-08-11/T671}}
  \end{picture}}
\newcommand{\identity}{\mathbf{1}}
\newcommand{\Tr}{\operatorname{Tr}} 

\newcommand{\Fig}[1]{Fig.~\ref{#1}}

\newcommand{\ndf}{\mathsf{ndf}}                     

\newcommand{\smallfrac}[2]{\mbox{\small ${\displaystyle
      \frac{#1}{#2}}$}}

\title{The strong-coupling limit of minimal lattice 
       \hbox{Landau gauge}\preprint
}

\ShortTitle{The strong-coupling limit of minimal lattice Landau
    gauge}

\author{\speaker{Andr\'e Sternbeck} and 
        Lorenz von Smekal\\
        Centre for the Subatomic Structure of Matter
        (CSSM), School of Chemistry \& Physics,\\
        The University of Adelaide, SA 5005, Australia\\
        \mbox{E-mail: \email{andre.sternbeck@adelaide.edu.au},
        \email{lorenz.smekal@adelaide.edu.au}}
}

\abstract{We study the gluon and ghost propagators of lattice Landau
  gauge in the strong coupling limit $\beta = 0$ in pure SU(2)
  lattice gauge theory to find evidence of the conformal infrared
  behaviour of these propagators as predicted by a variety of
  functional continuum methods for asymptotically small momenta $q^2
  \ll \Lambda_\mathrm{QCD}^2$. In the strong-coupling limit, this same
  behaviour is obtained for the larger values of $a^2q^2$ (in units of
  the lattice spacing $a$), where it is otherwise swamped by the gauge
  field dynamics. Deviations for $a^2 q^2 < 1 $ are well parametrized
  by a transverse gluon mass $\propto 1/a$. Perhaps unexpectedly,
  these deviations are thus no finite-volume effect but persist in the
  infinite-volume limit. They furthermore depend on the definition of
  gauge fields on the lattice, while the asymptotic conformal behaviour
  does not.}

\FullConference{The XXVI International Symposium on Lattice Field Theory\\
		 July 14-19 2008\\
		 Williamsburg, Virginia, USA}

\begin{document}

\section{Introduction}

The infrared behaviour of the QCD Green's functions contains essential
information about the realisation of confinement in the covariant
formulation of QCD, especially in Landau gauge
\cite{Alkofer:2000wg}. It has been focus of intensive research  
over the past ten or so years and was studied with a variety of
methods. These include different continuum-based functional methods 
\cite{vonSmekal:1997isvonSmekal:1997vx,Lerche:2002ep,
Zwanziger:2001kw,Pawlowski:2003hq} as well as
Monte Carlo (MC) simulations. In particular, the former have
predicted that, in Landau gauge, the deep-infrared behaviour of the
momentum-space gluon and ghost dressing  functions, $Z$ and $G$,
should be given by \cite{vonSmekal:1997isvonSmekal:1997vx} 
\begin{equation}
   \label{eq:infrared-gh_gl}
   \centering
   Z(p^2) ~\sim~ (p^2/\Lambda^2_\mathrm{QCD})^{2\kappa_Z}
   \quad\text{and}\quad
   G(p^2) ~\sim~ (p^2/\Lambda_\mathrm{QCD}^2)^{-\kappa_G} 
   \qquad\text{for}~~p^2 \to 0\,,
\end{equation}
which are both determined by a unique critical infrared
exponent $\kappa_Z = \kappa_G \equiv \kappa\in (0.5,1)$. 
Under a mild regularity assumption
on the ghost-gluon vertex \cite{Lerche:2002ep}, the value of this
exponent is furthermore obtained as \cite{Lerche:2002ep,Zwanziger:2001kw} 
\vspace{-.2cm}
\begin{equation} 
  \kappa \, = \, (93 - \sqrt{1201})/98 \, \approx \, 0.595\,.
\label{kappa_c}
\end{equation}
The prediction of this so-called conformal infrared behaviour has been
challenged by a number lattice simulations. In fact, with the notable
exception of pure  SU(2) lattice gauge theory in  2 dimensions
\cite{Maas:2007uv}, the proposed infrared behaviour
(\ref{eq:infrared-gh_gl}), or even its onset, has not
conclusively been confirmed in lattice simulations so far. In
contrast, numerical evidence is provided that the dressing functions are
consistent with a transverse gluon mass, $Z \sim
p^2/M^2 $ and $G = $ const., for $p^2 \to 0$.

Finite-volume effects were soon identified as the prime suspects 
responsible for the disagreement between continuum and lattice results
\cite{Fischer:2007pf}. In fact, in order to be able to see an onset of
an at least approximate conformal infrared behaviour in a finite volume
of extend $L$, a reasonably large range of momenta satisfying 
\vspace{-.2cm}
\begin{equation}
  \pi/L \, \ll p \, \ll \Lambda_{\mathrm{QCD}} 
\label{scales}
\end{equation}
needs to be accessible, and it was concluded in \cite{Fischer:2007pf}
that lattice sizes of about 15~fm are needed before even an onset of
the leading infrared behaviour can be observed. Not only this has
triggered a run for the biggest-ever lattice sizes used in simulations
(see, e.g.,
\cite{Sternbeck:2007ugBogolubsky:2007udCucchieri:2007mdCucchieri:2007rg}).
All those simulations, however, show very little if no tendency to follow the
finite-size corrections to the predicted asymptotic behaviour, even
though their volumes reached up to about 20~fm in size.

In this study, we focus on the gluon and ghost propagator in
the strong coupling limit, $\beta\to0$, of pure $SU(2)$ lattice Landau
gauge. This unphysical limit, which can be interpreted as the formal
limit $\Lambda_\mathrm{QCD} \to \infty $, allows us to assess whether
the predicted conformal behaviour can be seen for the larger lattice
momenta $p$, after the upper bound in (\ref{scales}) has been removed,
in a range where the dynamics due to the gauge action would otherwise
dominate and cover it up completely. Our complete analysis can be found
in \cite{fullpaper}.

\section{Gluon and ghost dressing functions in the strong coupling
  limit}

We simulate SU(2) gauge theory in the strong coupling limit by
generating random link configurations $\{U\}$. 
These are sets of SU(2) gauge links, $U_{x\mu} =
u^0_{x\mu}\identity + i\sigma^au^a_{x\mu}$, equally distributed
over~$(u^0,\vec{u})_{x\mu}\in S^3$. Those configurations are fixed to the
(standard lattice) Landau gauge (SLG) using an over-relaxation
algorithm that iteratively minimises the SU(2) gauge functional of
SLG, 

\begin{equation}
  V_U[g] = 4 \sum_{x,\mu}\left(1-\smallfrac{1}{2} \Tr
    U^g_{x\mu}\right) \; .
  \label{eq:functional}
\end{equation}
To satisfy Landau gauge with sufficient accuracy the over-relaxation
algorithm is iterated until the stopping criterion
~$
  \max_x\Tr\big[(\nabla^b_{\mu} A_{x\mu}^g)(\nabla^b_{\mu}
  {A^{g\,\dagger}_{x\mu}}) \big]< 10^{-13}
$~
is met at every lattice site. Here
$\nabla^b_{\mu}$ denotes the lattice backward derivative and
$A_{x\mu}^g$ is the lattice gluon field of SLG, given by 
\begin{equation}
  A_{x\mu}^g = \frac{1}{2ia}\left(U^g_{x\mu} -
    U^{g\,\dagger}_{x\mu}\right)
  \label{eq:gluonfield_SLG}
\end{equation}
in terms of the gauge-transformed link $U^g_{x\mu}$. Gluon and ghost
propagators are then calculated in momentum space employing standard
techniques. For further details refer to \cite{fullpaper}.

\begin{figure*}
  \centering
  \mbox{\includegraphics[height=6cm]{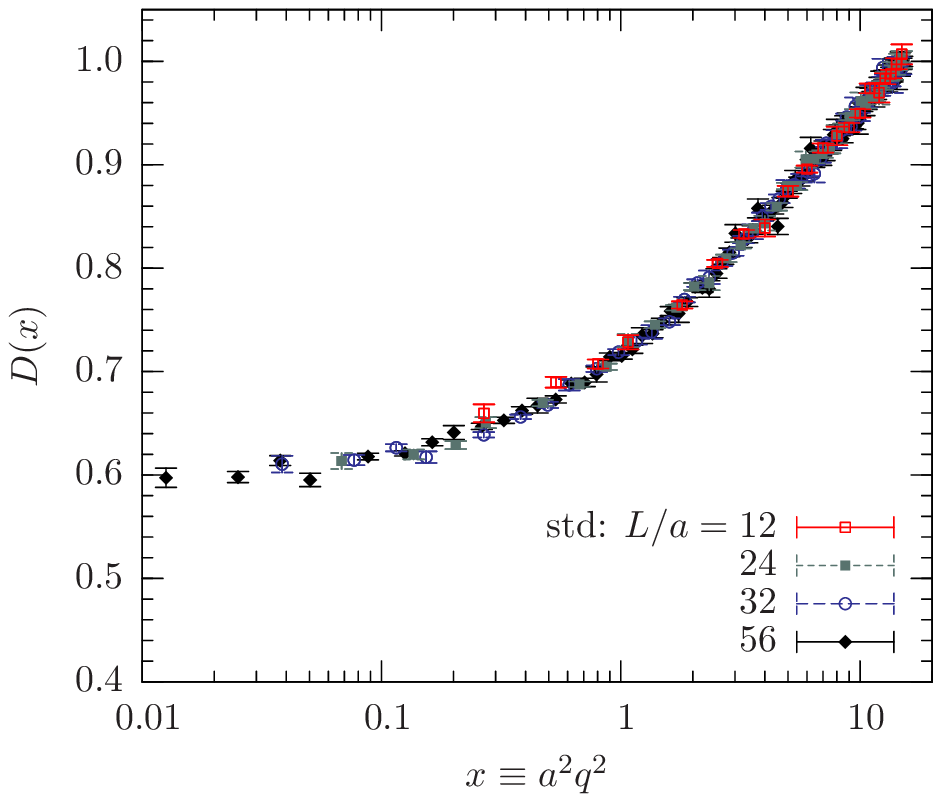}\qquad
        \includegraphics[height=6cm]{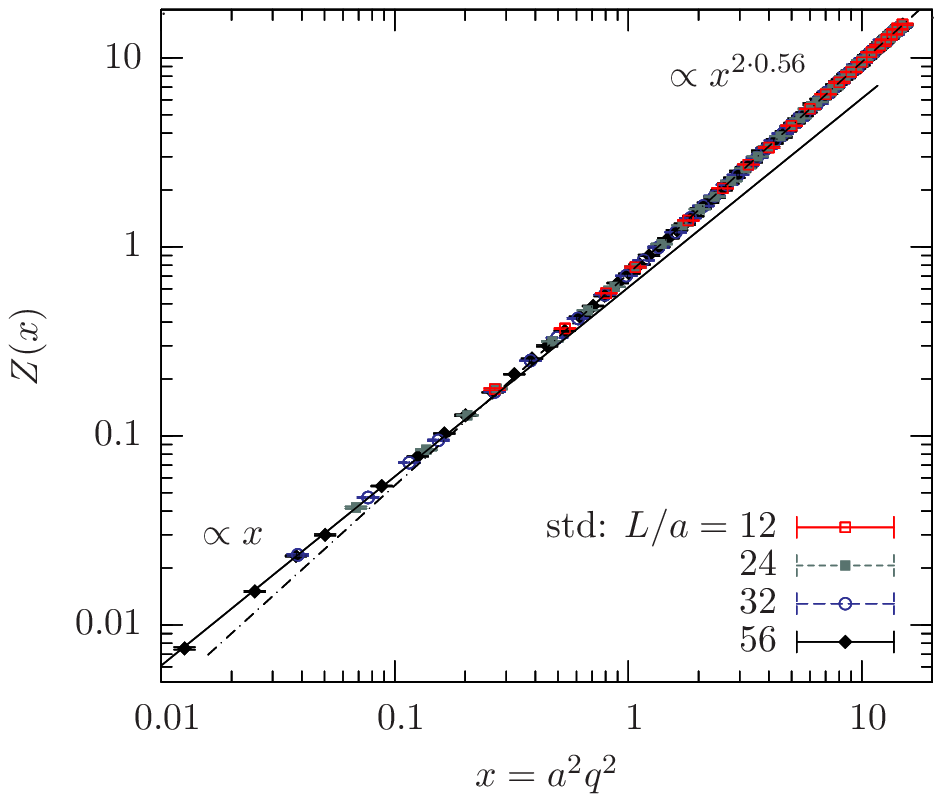}}
  \caption{The gluon propagator (left) and its dressing function 
    (right) vs.\ lattice momentum $a^2q^2$ for different lattice sizes in
    the strong coupling limit. Straight lines (right) represent fits of
    the $56^4$ data to its asymptotic behaviour at low (solid) and high
    (dashed) momenta (see \cite{fullpaper}).} 
  \label{fig:gl_dress_qq_beta0-stdLG-U-Ud}
\end{figure*}

The data for the gluon propagator and its dressing function is shown in
\Fig{fig:gl_dress_qq_beta0-stdLG-U-Ud} versus the lattice momentum
$a^2q^2$. The propagator is seen to increase with momentum, while it
plateaus at low momenta. Perhaps unexpectedly, however, this happens
irrespective of the lattice size at around $a^2 q^2 \approx 1$. It is
therefore not a finite-volume effect, and the observed mass behaves as 
$D^{-1}(0) \propto M^2 \propto 1/a^2$ in the strong-coupling limit
with no significant dependence on $L$.  

The asymptotic behaviour at either end is also seen in the
corresponding dressing function $Z$. We compare this function to fits
obtained from the $56^4$ data in the right panel of
\Fig{fig:gl_dress_qq_beta0-stdLG-U-Ud}: Writing $x\equiv a^2q^2$, we
use $f(x) = c x^{2\kappa_Z}$ as the fitting function for $x\in[2,14]$
and $f(x) = d x$ for small momenta $x<0.1$ to fit $c$, $d$ and the
exponent $\kappa_Z$.  The data at large momenta is well described by
the power law (\ref{eq:infrared-gh_gl}) with an exponent $\kappa_Z =
0.563(1)$, while it approaches the massive behaviour at small $x$.

To assess the model dependence of these results we also used different
fit models and a broad range of fitting windows (see \cite{fullpaper}
for details).  Monitoring the corresponding $\chi^2/\ndf$ values we
find that
\vspace{-.2cm} 
\begin{equation}
  \label{eq:gl_fits}
  D_I(x) = cx^{2\kappa_Z-1}\qquad\text{and}\qquad 
  D_{II}(x) = c(d+x)^{2\kappa_Z-1}
\end{equation}
provide stable fits with similar values for $\kappa_Z$ and very little
dependence on the lattice size.  While $D_I(x)$ performs slightly
better at large momenta, $D_{II}(x)$ provides good stable fits of the
gluon propagator over the whole momentum range.

\clearpage

\begin{floatingfigure}[r]
    \includegraphics[width=0.55\linewidth]{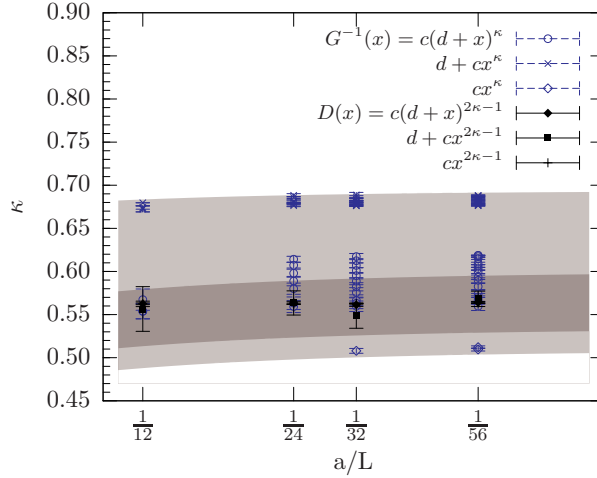}
  \caption{$\kappa$ versus $a/L$ for the gluon and ghost propagators.} 
  \label{fig:kappa_gl_gh_L_fits}
\end{floatingfigure}
The $\kappa_Z$ values extracted from different fit models and lattice
sizes with fixed window $x\in[2,14]$ are shown in
\Fig{fig:kappa_gl_gh_L_fits}. The observed dependencies
on either one are rather small.  There is a general trend for
$\kappa_Z$ to slightly increase with $a/L$ (the dark grey band in
\Fig{fig:kappa_gl_gh_L_fits}) though this is within the systematic
uncertainty due to the fit model.

Similar fits were performed to extract the exponent $\kappa_G$ from
the ghost dressing function $G$. These fits are less robust with a
more pronounced model dependence (light grey band in
\Fig{fig:kappa_gl_gh_L_fits}). This is mainly due to the wider
transition region, from $G = $ const.~at small $x$ to $G\sim
x^{-\kappa_G} $ at large $x$, which is under less control here.  
The exponent can nevertheless be estimated as $\kappa_G = 0.60(7)$. 
The results are consistent with the scaling relation $\kappa_Z = \kappa_G$.

\section{Comparing different lattice definitions of gauge fields} 
\label{sec:comparing_diff_latt_def}

It is interesting to compare the above presented SLG data to that of
the \emph{modified lattice Landau gauge} (MLG) introduced in
\cite{vonSmekal:2007ns} which is based on stereographic projection to
define lattice gauge fields. When comparing MLG to the ever popular
SLG, there is no advantage that the SLG has over the MLG. A promising
particular feature of the MLG on the other hand is that it provides a
way to perform gauge-fixed MC simulations sampling \emph{all} Gribov
copies of either sign (of the Faddeev-Popov determinant) in the spirit
of BRST. This feature will be explored in a forthcoming study.  Here
we simply use the MLG for comparison in the standard way, {\it i.e.},
we gauge-fix configurations via minimisation of the MLG functional,
which in SU(2) is of the form
\cite{vonSmekal:2007ns}
\begin{equation}
  \label{eq:functional_modLG}
  \widetilde V_U[g] = -8 \sum_{x,\mu}
  \ln\left(\frac{1}{2}+\frac{1}{4}\Tr 
    U^{g}_{x,\hat{\mu}} \right)\;. 
\end{equation}
The MLG functional differs from that of SLG (\ref{eq:functional}) and so
do the lattice gluon fields of MLG,
\begin{equation}
  \widetilde{A}_{x\mu} = \frac{1}{2ia}\left(\widetilde{U}_{x\mu} -
    \widetilde{U}^{\dagger}_{x\mu}\right)
  \qquad\text{where}\quad
  \widetilde{U}_{x\mu} 
     \equiv \frac{2 U_{x\mu}}{1+ \frac{1}{2} \Tr U_{x\mu}} \; . 
  \label{eq:gluonfield_MLG}
\end{equation}
The lattice Landau gauge condition $F_x = \nabla^b_\mu A_{x\mu} = 0$
is formally unchanged in the stereographically projected gauge fields
$\widetilde A$, in particular, we have $\widetilde F_x(A) =
F_x(\widetilde A)$. The corresponding Faddeev-Popov operator is given
explicitly in \cite{vonSmekal:2007ns}.  Both lattice definitions of
Landau gauge have the same continuum limit, and any differences
between MLG and SLG data at finite lattice spacings are lattice
artifacts. It is also worth mentioning that gauge configurations fixed
to MLG do not satisfy the gauge condition of SLG and vice
versa. Nonetheless, exact transversality, {\it i.e.},\ $q_{\mu}(k)
A_{\mu}(k) = 0$, is satisfied at finite lattice spacing $a$ for both
of them equally, if the momenta $q_\mu(k)$ are defined as
~$aq_{\mu}(k) := 2\sin(\pi k_{\mu}/N_{\mu})$ with integer valued
$k_{\mu}\in(-N_{\mu}/2,N_{\mu}/2]$ in the usual way.

\begin{figure*}[b]
    \includegraphics[height=5.7cm]{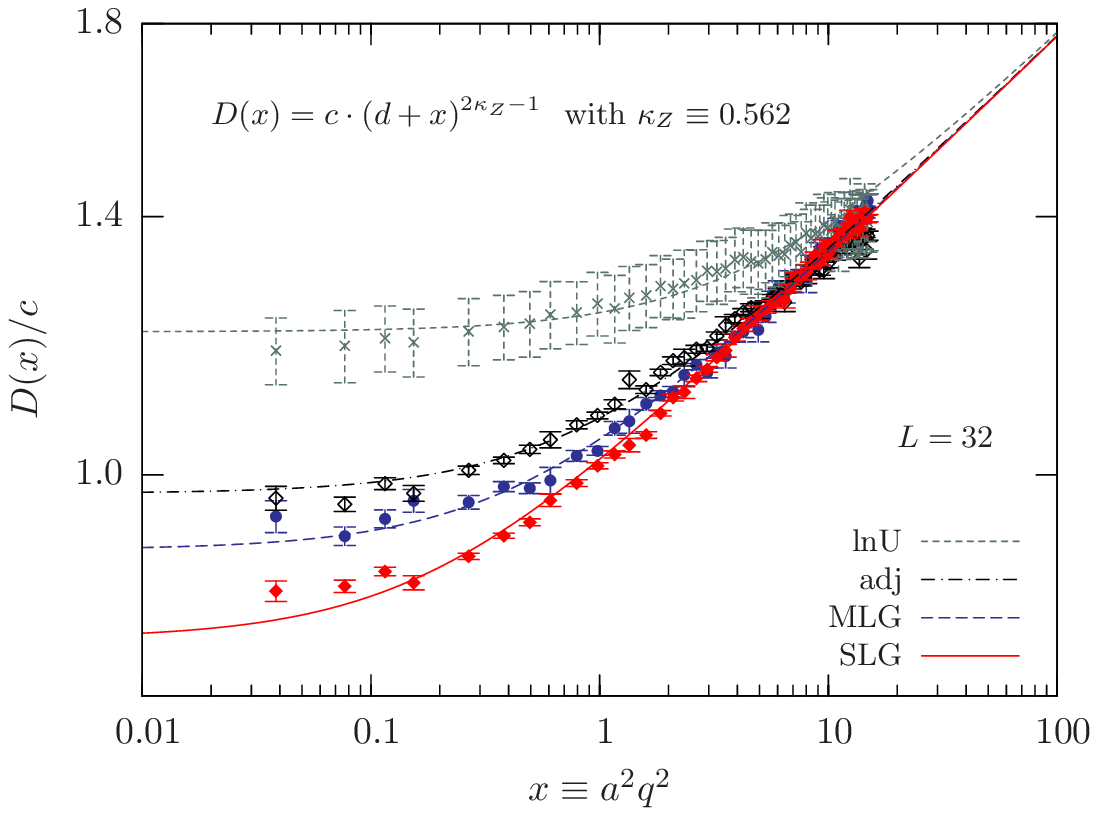}
    \includegraphics[height=5.7cm]{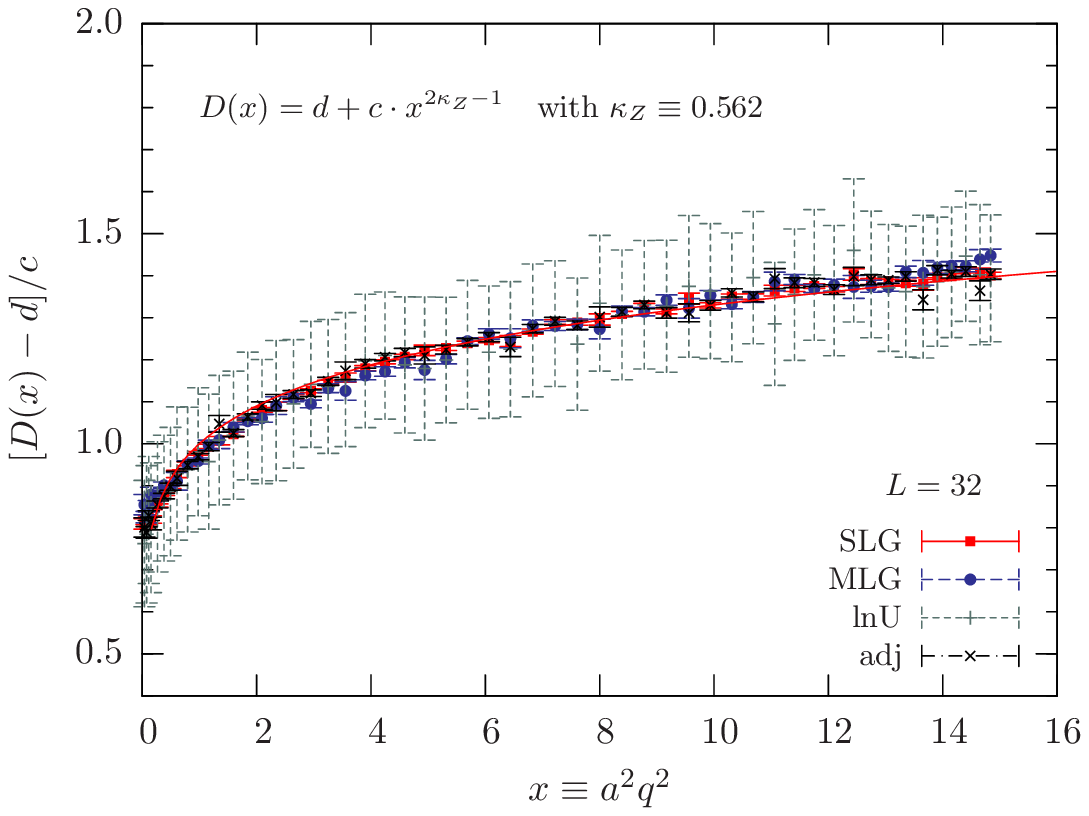}
    \caption{The gluon propagator in the strong-coupling limit over 
      $x=a^2q^2$ for the various definitions of gauge fields mentioned
      in the text. All data from $32^4$ lattices and normalised to the
      scaling branch. All fits were performed with fixed
      $\kappa_Z=0.562$ as obtained for the SLG (with $L/a=32$). On the
      left, the results are shown for $D_{II}(x)$ in (\protect\ref{eq:gl_fits}). 
      On the right we use $D_{III}=d+cx^{2\kappa_Z-1}$ to demonstrate
      how the data from all definitions collapse to a unique curve
      $\propto x^{2\kappa_Z-1}$ (solid, red) when the mass term is
      subtracted.} 
    \label{fig:gl_qq_beta0_global}
\end{figure*}

The data for the gluon propagator of SLG (red filled diamonds) is
compared to that of MLG (blue filled circles) in
\Fig{fig:gl_qq_beta0_global}. There we also show data for the gluon
propagator where either
\begin{equation}
  \label{eq:lnUdef_adjdef}
    a A^{\mathsf{adj}}_{x\mu} \,=\,  u^{0}_{x\mu} u^{a}_{x\mu}
    \sigma^a    \; ,   \qquad\textrm{or}\qquad
    a A^{\mathsf{ln}}_{x\mu} \,=\, 
    \phi_{x\mu}^a \sigma^a/2 \quad \textrm{from} \;\;  U_{x\mu} =
    \exp\{ i \phi_{x\mu}^a \sigma^a/2 \} 
\end{equation}
are used to define lattice gluon fields based on the adjoint
representation, $A^{\mathsf{adj}}$ (black open diamonds), and thus
blind to the centre \cite{Langfeld:2001cz}, or on the tangent space at
the identity $A^{\mathsf{ln}}$ (green crosses).  In these two cases,
$A^{\mathsf{adj}}$ and $A^{\mathsf{ln}}$, for the purpose of a
qualitative comparison, we simply use the gauge configurations of the
SLG to calculate the gluon propagator.  Especially for
$A^{\mathsf{ln}}$ this implies, however, that the condition
$q_{\mu}(k)A_{\mu}(k)=0$ is satisfied at best approximately and
nowhere near the precision of the other two (SLG and MLG). The
residual uncertainty due to other possible tensor structures then
causes the somewhat larger errors for this definition as seen in
\Fig{fig:gl_qq_beta0_global}.

First, we fit the data from all four definitions to $D_{II}(x)$ in
(\ref{eq:gl_fits}) which provides the best overall description in the full
momentum range. In order to demonstrate how the other definitions compare to 
the SLG, we keep its value for the exponent fixed when fitting
the other data, {\it i.e.}, $\kappa_Z = 0.562$ as obtained for $L/a =
32$ in SLG is used in all fits. Relative to the scaling branch
$\propto x^{2\kappa_Z-1}$ for  large $x=a^2q^2 $ we then observe a
strong definition dependence in the (transverse) gluon mass term at
small $x$ (see the left panel in \Fig{fig:gl_qq_beta0_global}).
The relative weight of the two asymptotic branches,
scaling at large $x$ and massive at small, is clearly discretisation 
dependent and can not be compensated by finite renormalisations. 
A first indication that the massive branch might indeed be the
ambiguous one is the observed  $M\propto 1/a$. This is consistent with the
fact that the definitions of gauge fields on the lattice, which agree
at leading order, all differ at order $a^2$, and so do their
corresponding Jacobian factors which leads to lattice mass
counter-terms of different strengths. The observation that these
differences matter here explicitly demonstrates the breakdown of the
lattice Slavnov-Taylor identities in minimal lattice Landau gauge in
the non-perturbative domain. 

To assess whether this ambiguity has an influence on the exponent $\kappa_Z$,
we have also used fits of the form $D_{III}=d+cx^{2\kappa_Z-1}$, again
with the fixed $\kappa_Z\equiv0.562$ from the $32^4$ SLG data. This
form leads to somewhat larger $\chi^2/\ndf$ that arise from the transition
region around $x=a^2q^2 \sim 1$ which is described a bit better by $D_{II}$.
If we then subtract the constant $d$, however, the normalised data of
all four definitions nicely collapse onto a unique curve $\propto
x^{2\kappa_Z-1}$ as seen in the right panel of
\Fig{fig:gl_qq_beta0_global}. The data is fully consistent with a
unique exponent $\kappa_Z $ of around the SLG value which has an
infinite volume extrapolation of $0.57(3)$.

The strong-coupling ghost propagators of SLG and MLG are presented in
\cite{fullpaper}. The best global fits are then obtained for the form
$G^{-1}(x) = d + c x^{\kappa_G}$. They are again consistent with a
unique scaling exponent $\kappa_G \approx \kappa_Z$ and deviate from
one another in the relative strength of the constant term at small
$x=a^2q^2$.  Instead of discussing the strong-coupling ghost
propagator in more detail here we turn directly to the product of
dressing functions defining a running coupling.

\section{Coupling constant}
\label{sec:coupling}

The predicted infrared scaling (\ref{eq:infrared-gh_gl}) with
$\kappa_Z = \kappa_G$ immediately implies that the running coupling
defined by \cite{vonSmekal:1997isvonSmekal:1997vx}
\vspace{-.2cm}
\begin{equation} 
 \alpha_s(p^2) \, = \, \frac{g^2}{4\pi} Z(p^2) G^2(p^2)
\label{alpha_minimom}
\end{equation}

\begin{floatingfigure}[r]
  \centering
  \includegraphics[width=0.52\linewidth]{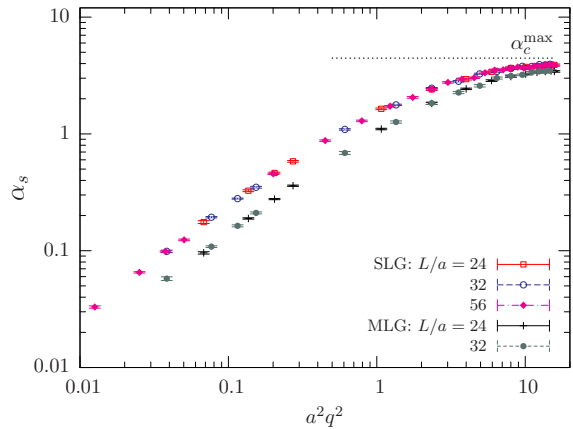}
  \caption{The strong-coupling limit of $\alpha_s$ from
    (\ref{alpha_minimom}) in minimal lattice Landau gauge
     for SLG and MLG. The dotted line marks  $\alpha^{\mathrm{max}}_c
     \approx 4.46$, the maximal value of the critical coupling
     for $SU(2)$ from the continuum prediction \cite{Lerche:2002ep}.}
  \label{fig:alpha_qq_beta0}
\end{floatingfigure}
\noindent
approaches an infrared fixed point, $\alpha_s \to \alpha_c $ for $p^2
\to 0 $. Standard continuum conventions of course need rescaling
$g^2 Z \to Z$  when comparing to lattice definitions such as
(\ref{eq:gluonfield_SLG}). The predicted conformal scaling in the
strong-coupling limit, with $Z = c_Z (a^2q^2)^{2\kappa} $ and $G^{-1} = c_G
(a^2q^2)^\kappa$, would therefore imply that the coupling 
(\ref{alpha_minimom}) is indeed constant with  $\alpha_s = \alpha_c =
c_Z/(4\pi c_G^2)$. Note that its value is thus determined precisely by those
multiplicative constants in the propagators that have been
ignored in the analysis up to here. They have to be extracted from
the  bare lattice data without rescaling or renormalisation.
In complete agreement with the general observation of conformal 
scaling in the strong-coupling limit at large momenta the product
(\ref{alpha_minimom}) of the gluon and ghost dressing functions levels
at $\alpha_c \approx 4$ for large $a^2q^2$. As expected for an
exponent $\kappa$ slightly smaller than the  value in (\ref{kappa_c})
this is just below the upper bound $\alpha^{\mathrm{max}}_c \approx
4.46$ for $SU(2)$, see \cite{Lerche:2002ep}. It is quite compelling
that also this result is nearly independent of the gauge-field
definition, {\it i.e.}, almost identical for SLG and MLG, see
\Fig{fig:alpha_qq_beta0}. Predominantly driven by the ghost
propagator, the violations to this conformal infrared scaling set
in as soon as the ambiguity in the definition of minimal lattice
Landau gauge does.

\clearpage 

\section{Conclusions}

We have studied gluon and ghost propagators of pure SU(2) minimal
lattice Landau gauge in the strong coupling limit. This unphysical
limit probes the gauge field measure of the minimal lattice Landau
gauge for there is no contribution from the Yang-Mills (plaquette)
action.

As expected for a formal limit $ \Lambda_\mathrm{QCD} \to \infty$, it
is then observed that the propagators show a conformal scaling
behaviour (\ref{eq:infrared-gh_gl}) for large lattice momenta, $a^2
q^2 \gg 1$. Finite-size effects are small and the combined gluon and
ghost data is consistent with an $L/a \to \infty $ extrapolation of a
critical exponent $\kappa = 0.57(3)$.  This scaling branch at large
$a^2q^2 $ furthermore leads to a critical coupling of $\alpha_c
\approx 4$ which is just below the predicted maximum
$\alpha^{\mathrm{max}}_c \approx 4.46$ for $SU(2)$. These results show
very little if no significant dependence on the lattice definition of
gauge fields and measure.

Another unambiguous result is the emergence of a transverse gluon mass
$M\propto 1/a$ in the strong-coupling limit of minimal lattice Landau
gauge. Both propagators show this massive behaviour at small momenta
corresponding to $Z \sim q^2/M^2 $ and $G = $ const. for $a^2q^2 \ll
1$. This massive low momentum branch of the data, however, depends
strongly on which lattice definition is being used for the gauge
fields and their measure. This is typical for a mass counter-term on
the lattice and demonstrates the breakdown of lattice Slavnov-Taylor
identities (STIs) and BRST symmetry in minimal lattice Landau gauge
beyond perturbation theory. It is still possible that this ambiguity
disappears in the continuum limit, eventually. But because it is a
combination of ultraviolet (mass counter-term) and infrared (breakdown
of STIs) effects, this might take very fine lattice spacings in
combination with very large volumes and therefore who-knows-how big
lattices to verify explicitly.


\medskip
\noindent
{\small This research was supported by the Australian Research Council and by
eResearch South Australia.}


\providecommand{\href}[2]{#2}\begingroup\raggedright\endgroup

\end{document}